\documentclass[a4paper,11pt]{article}
\usepackage{cite}
\usepackage{amsmath,amssymb,amsfonts}
\usepackage{breqn}
\usepackage{algorithmic}
\usepackage{graphicx}
\usepackage{textcomp}
\usepackage{xcolor}
\usepackage{graphics}
\usepackage{multirow}
\usepackage{balance}
\usepackage{url}
\def\BibTeX{{\rm B\kern-.05em{\sc i\kern-.025em b}\kern-.08em
    T\kern-.1667em\lower.7ex\hbox{E}\kern-.125emX}}

\begin{document}
\title{Deep auscultation:\\ Predicting respiratory anomalies and diseases via recurrent neural networks}


\author{Diego Perna\\
\textit{DIMES} - \textit{University of Calabria}, Rende (CS), Italy \\
d.perna@dimes.unical.it
\and
Andrea Tagarelli\\
\textit{DIMES} -  \textit{University of Calabria}, Rende (CS), Italy \\
andrea.tagarelli@unical.it}

\date{}

\maketitle
\begin{abstract}
Respiratory diseases are among the most common causes of severe illness and death worldwide. Prevention and early diagnosis are essential to limit or even reverse the trend  that characterizes the diffusion of such diseases. In this regard, the development of advanced computational tools for the analysis of respiratory auscultation sounds can become a game changer for detecting disease-related anomalies, or diseases themselves. In this work, we propose a novel learning framework for respiratory auscultation sound data. Our approach combines state-of-the-art feature extraction techniques and advanced deep-neural-network architectures. Remarkably, to the best of our knowledge, we are the first to model a recurrent-neural-network based learning framework to support the clinician in detecting  respiratory diseases, at either level of  abnormal  sounds or  pathology classes. Results obtained on the ICBHI benchmark dataset show that our approach outperforms competing methods on both anomaly-driven and pathology-driven prediction tasks, thus advancing the state-of-the-art in respiratory disease analysis. 
\end{abstract}


\section{Introduction}
%

With the term ``the Big Five'', the World Health Organization  identifies five respiratory diseases among the most common causes of severe illness and death worldwide, namely chronic obstructive pulmonary disease (COPD), asthma, acute lower respiratory tract infection (LRTI), tuberculosis, and lung cancer~\cite{who}. 
%
%
 %
  The number of people affected by COPD reaches 65 million, with about 3 million deaths per year, making it the third leading cause of death worldwide~\cite{cruz2007global,burney2015global}. 
Asthma is a common chronic disease that is estimated to affect as many as 339 million people worldwide~\cite{network2018global}, and it is considered the most common chronic childhood disease. 
 %
    Another widespread disease 
which especially affects children under 5 years old is pneumonia~\cite{wardlaw2006pneumonia}. The \textit{Mycobacterium tuberculosis}   agent has infected over 10 million people, and it is considered the most common lethal infectious disease~\cite{tb2016world}. 
Yet, lung cancers  kill  around 1.6 million people every year~\cite{cancerstats}.  

Prevention, early diagnosis, and treatment are     key factors to limit the spread of such diseases and their negative impact on the length and quality of life. 
Lung auscultation is an essential part of the respiratory examination  and is helpful in diagnosing various disorders, such as  \textit{anomalies} that may occur in the form of abnormal sounds (e.g., \textit{crackles} and \textit{wheezes}) in the respiratory cycle. 
When performed through   advanced computational methods, a deep analysis of such sounds can be of great support to  the physician, which could result in enhanced detection of respiratory diseases. 
 %
 %
 
In this context,  \textit{machine learning} techniques have shown to provide an invaluable computational tool for detecting disease-related anomalies in the early stages of a respiratory dysfunctions (e.g.,~\cite{berouti1979enhancement,serbes2018automated,kochetov2018noise}). 
%
%
 In particular,   \textit{deep learning} (DL) based methods promise to support enhanced detection of respiratory diseases from auscultation sound data, given their well-recognized ability of    
  learning complex non-linear functions from large, high-dimensional data.      In recent years, this has led DL methods to set  state-of-the-art performances in a wide range of domains, such as machine translation,  image segmentation,   speech and signal  recognition.

In this work, we aim to advance the state-of-the-art in research on machine-learning detection of respiratory anomalies and diseases through the use of advanced DL architectures. A major contribution of our work  is the definition of a learning framework based on \textit{Recurrent Neural Network}  (RNNs) models to effectively handle respiratory disease prediction problems at both anomaly- and pathology-levels.  
 Unlike other types of DL networks, RNNs are designed to effectively discover the time-dependent patterns from sound data. 
 To the best of our knowledge, the use of such models to address the above problems  has not been adequately studied so far. 
 We also contribute with a preprocessing methodology for a flexible extraction of core groups of cepstral features to feed the inputs to an RNN model. 
  Remarkably, our RNN models  were trained and tested using the \textit{ICBHI Challenge} dataset, which provides an unprecedented,  reproducible and standardized benchmark on which new algorithms can be fairly evaluated and compared~\cite{data}.  
  Results obtained on the ICBHI benchmark, according to different assessment criteria, highlight the superiority of our RNN-based methods against all selected competitors that participated to the ICBHI Challenge, as well as against a further competitor based on a DL framework.  
  

\section{The ICBHI Challenge}
\label{sec:data}


The ICBHI Challenge dataset~\cite{data}    was built in the context of a challenge on respiratory data analysis organized in conjunction with the 2017 Int. Conf. on Biomedical Health Informatics (ICBHI). The dataset contains audio samples that were collected independently by   two research teams in two different countries. 
 The data acquisition process was characterized by varying  recording equipment, microphone chest position, environmental noise, 
  etc. Such variability raised the level of difficulty of the challenge by introducing several sources of noise and unpredictability.  
%

{\bf Annotations.\ }The ICBHI sound data were provided with two types of annotation: i) for each respiratory cycle, whether or not  crackles and/or wheezes are present, and ii) for every patient,  whether or not  a specific pathology from a set of predetermined categories is present. 
 As we shall discuss in Sect.~\ref{sec:related}, all the participants to the ICBHI Challenge  focused on the first, finer-grain type of annotations. 
To advance research on respiratory data analysis, in this work we also take the opportunity of  exploiting the ICBHI Challenge to assess and comparatively evaluate our proposed framework on prediction tasks at either level of anomalies and pathologies. 



\subsection{Abnormal sounds}
\label{sec:crackles_and_wheezes}
Crackles and wheezes are commonly referred to by   domain experts as criteria to assess the  health status of  a patient's respiratory system. 
 %
 %
 We adopt the definitions provided by  The European Respiratory Society (ERS)   on Respiratory Sounds and described in~\cite{pasterkamp2016towards}. 

Crackles are discontinuous, explosive, and non-musical adventitious lung sounds, which are   
 usually classified as \textit{fine} or \textit{coarse} crackles based on their duration, loudness, pitch, timing in the respiratory cycle, and relation  to coughing and changing body position.  
The two types of crackles are normally distinguished based on their  duration:      longer than 10~ms for coarse crackles, and shorter than 10~ms  for fine crackles. The frequency range of   crackles   is 60-2000~Hz, with most informative frequencies up to  1200~Hz~\cite{sarkar2015auscultation}.

Conversely, wheezes are high-pitched continuous, musical, and adventitious lung sounds, usually characterized by a dominant frequency of 400~Hz (or higher) and sinusoidal waveforms.
Although the standard definition of continuous sound includes a duration longer than 250~ms, a wheeze does not necessarily extend beyond 250~ms  and is usually   longer than 80-100~ms.
Severe obstruction of the intrathoracic lower airway or upper airway obstruction can be associated with inspiratory wheezes. Asthma and chronic obstructive pulmonary diseases (COPD) patients develop generalized airway obstruction. However, wheezing  could even be detected in a healthy person towards the end of expiration after forceful expirations~\cite{sarkar2015auscultation}.

\subsection{Respiratory data}
The ICBHI Challenge database consists of a total of 5.5 hours of recordings containing 6898 respiratory cycles, of which 1864 contain crackles, 886 contain wheezes, and 506 contain both crackles and wheezes, in 920 annotated audio samples from 126 subjects.

\begin{figure}[t!]
    \centering
    \includegraphics[width=\linewidth]{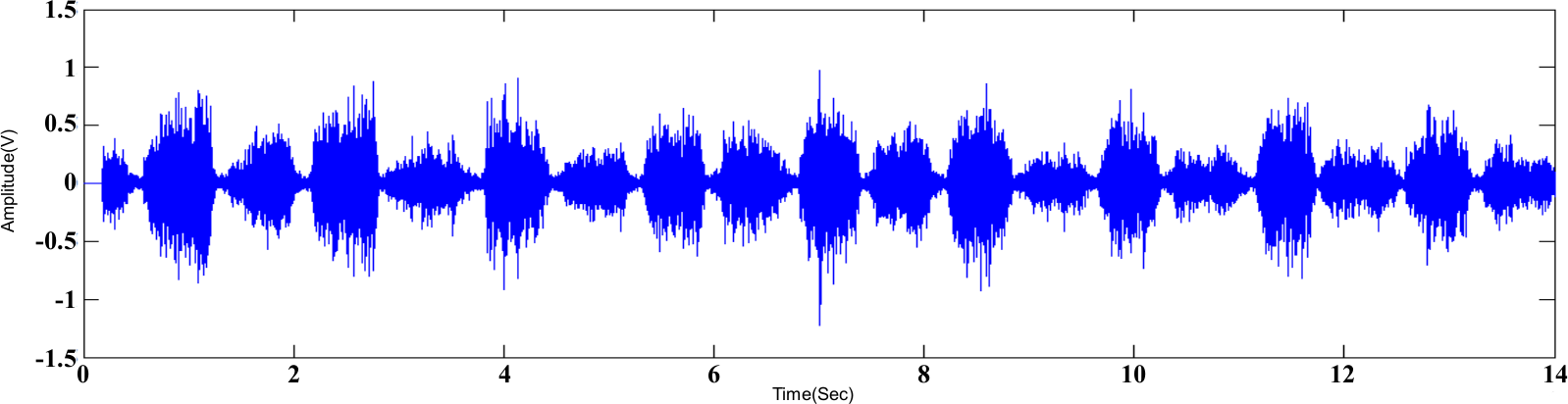}
    \caption{Example respiratory cycle waveform of a healthy patient.}
    \label{fig:resp_cycle}
\end{figure}

A single-channel respiratory sound, like the one shown in Figure~\ref{fig:resp_cycle}, 
 is composed of a certain number of cycles, which in turn include  four main components, two pauses, and two distinctive patterns. 
 Discarding fine-grain variations, mostly due to the conversion of air vibrations to electrical signal, a respiratory cycle is conventionally described as follows:  it starts from the inspiratory phase, 
   which  is characterized by a lower amplitude and a regular pattern, then it follows with an expiratory phase, 
   which  shows one or multiple peaks, a decreasing amplitude pattern, and   is usually characterized by a higher average energy.

As previously mentioned, the respiratory cycles were annotated by domain experts to  state the presence of crackles, wheezes, a combination of them, or no adventitious respiratory sounds. More in detail, the annotation style format includes the beginning of the respiratory cycle(s), as well as the end of the respiratory cycle(s), the presence or absence of crackles, 
 and the presence or absence of wheezes. 
The recordings were collected using heterogeneous equipment, with  duration ranging from 10~s to 90~s. The average duration of a respiratory cycle is 2.7~s, with a standard deviation of about 1.17~s; the median duration is about 2.54~s, whereas  the duration ranges from 0.2~s to above  16~s. 
  Moreover, wheezes are characterized by an average duration of about 600~ms, with a relatively high variance, and a minimum and maximum duration value ranging between 26~ms and 
   19~s; conversely, crackles are characterized by an average duration of about 50~ms,  smaller variance, and a minimum and maximum duration values of   3~ms and 4.88~s, respectively. 






\section{Related Work}
\label{sec:related}

We organize our discussion of  related work  into two parts, namely 
anomaly-driven prediction and pathology-driven prediction methods, depending on the target of classification of patients affected by respiratory diseases.  


\textbf{Anomaly-driven prediction.\ }
\label{sec:related:ICBHI}
In \cite{berouti1979enhancement}, the authors  proposed a method based on hidden Markov models  and Gaussian mixture models. The preprocessing phase includes a noise suppression step which relies on spectral subtraction~\cite{berouti1979enhancement}. The input of the model consists of Mel-frequency cepstral coefficients (MFCCs)   extracted in the range between 50~Hz and 2,000~Hz in combination with their first derivatives.  The method achieves  performance results up to  39.37\%, in compliance with the ICBHI score defined in \cite{jakovljevic2018hidden}. The authors also  tested an ensemble of 28 classifiers applying majority voting;  this approach led to  a 
slight  improvement of the performance of a single classifier, though   at the expense of ten times greater computational burden.

A method based on standard signal-processing techniques is described in \cite{serbes2018automated}. The   preprocessing phase here consists of a band-pass filter which is in charge of removing undesired frequencies due to heart sounds and other noise components. Then, the recording segment is separated into three channels, crackle, wheeze, and background noise, through resonance-based decomposition~\cite{selesnick2011wavelet}.  Subsequently,   time-frequency and time-scale features are extracted by applying short-time Fourier transform to each individual channel. The resulting features are finally aggregated and fed into a support vector machine classifier. This method achieves 49.86\% accuracy and an ICBHI score up to 69.27\%.

The MNRNN method proposed in \cite{kochetov2018noise} is designed to perform end-to-end classification with  minimal preprocessing needs.
MNRNN consists of three main components: i) a noise classifier based on two-stacked recurrent neural networks  which predicts noise label for every input frame, ii) an anomaly classifier, 
 and iii) a mask mechanism which is in charge of selecting only noiseless frames to feed into the anomaly classifier. MNRNN achieves 85\% accuracy in the detection of noisy frames, and   ICBHI score of 65\%.

The boosted decision tree model proposed in \cite{chambres2018automatic} utilizes two different types of features: MFCCs and low-level features extracted with the help of the \textit{Essentia} library~\cite{bogdanov2013essentia}.  
This method was mainly evaluated on a binary prediction setting  (i.e., healthy or unhealthy),   achieving  accuracy up to 85\%.

\textbf{Pathology-driven prediction.\ }
Differently from the above-mentioned methods, in our earlier work~\cite{perna2018convolutional}   we focused on   the prediction task from the perspective of  the pathology  affecting the patient. 
 Another key difference regards the input unit from which the coefficients have been extracted, which corresponds to a whole recording, rather than a respiratory cycle.   
The method in~\cite{perna2018convolutional}   is based on Convolutional Neural Networks (CNNs) and MFCCs coefficients, and exploits the class imbalance technique SMOTE. 

In this work, we tackle the anomaly-driven prediction problem, as well as the more challenging pathology-driven one.
Similarly to \cite{kochetov2018noise}, we define our method upon recurrent neural networks, but differently from it, we exploit the whole ICBHI dataset without omitting frames characterized by a high level of noise. 
In addition, like \cite{berouti1979enhancement, chambres2018automatic,perna2018convolutional}, our method also relies on MFCCs for the extraction of significant features from the respiratory sounds;  however, the use of an RNN architecture allows our model to benefit from the discovery of  time-dependent patterns, which otherwise would be ignored.


\section{Our Proposed Learning Framework}

\begin{figure*}[t!]
    \centering
    \includegraphics[width=1\linewidth]{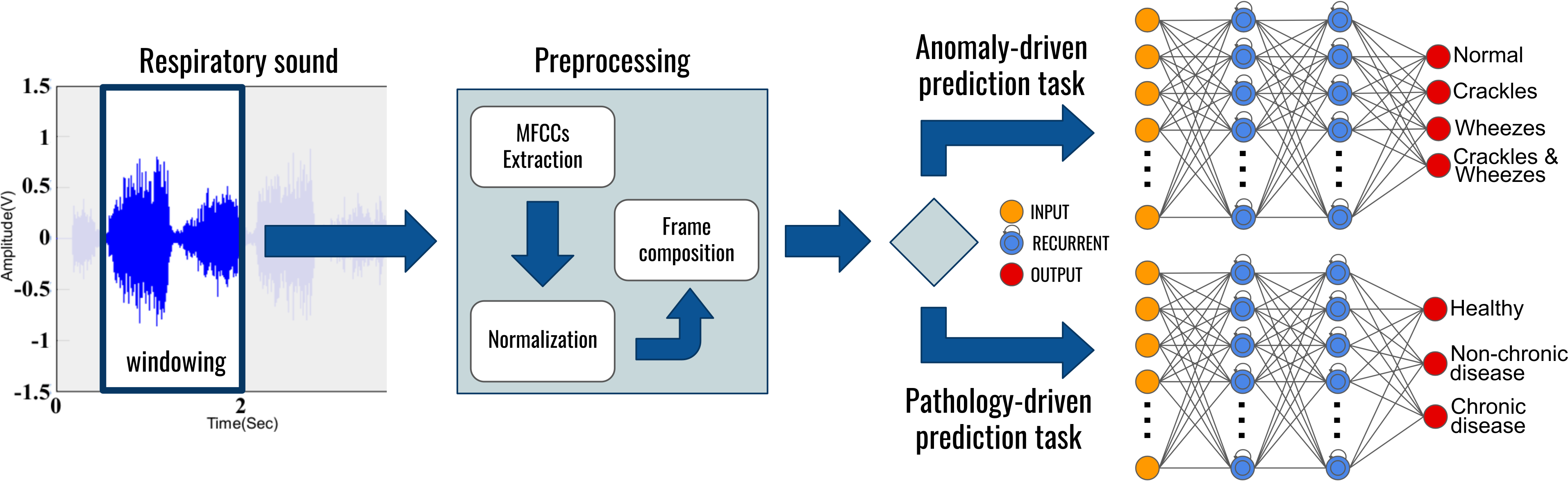}
    \caption{Illustration of our RNN-based framework for the prediction of respiratory anomalies and pathologies.}
    \label{fig:my_label}
\end{figure*}

In this section, we propose a novel   framework which leverages on a particularly suitable type of deep neural network architecture, namely \textit{recurrent neural networks} (RNNs).  Unlike existing approaches, our framework is designed to handle  a respiratory-disease prediction task at  anomaly-level (\emph{crackles} and \emph{wheezes}) or at pathology-level   --- \emph{chronic} diseases (COPD, bronchiectasis, asthma) and \emph{non-chronic} diseases (Upper and Lower Respiratory Tract Infection (URTI and LRTI), pneumonia, and bronchiolitis) 
--- at different resolutions (i.e., \emph{two-class} or \emph{multi-class} problems).  
Figure~\ref{fig:my_label} provides a schematic illustration of the workflow of our framework. 
In the following, we motivate and describe the use of RNNs, then we discuss in detail the preprocessing phase, and the criteria used in our evaluation.  

\subsection{Recurrent Neural Networks}
\label{subsec:dnn}
Traditional neural network architectures are based on the assumption that all inputs are sequentially independent. However, for many tasks, such as time-series analysis or natural language processing, in which the relations between consecutive training instances play a key role, this assumption is incorrect and could even be detrimental. 

The basic idea behind  RNNs  is to enable a network to remember past data with the goal of developing better models by leveraging sequential information~\cite{deeplearning}. 
The term ``recurrent'' suggests that this type of architecture is characterized by repeatedly performing the same action to the input sequence. However, the key distinguishing feature of RNNs is that the output depends on the current input as well as on the previously processed samples. The ability of  combining the informative content of the \textit{i-th} sample and the previously processed ones can be ascribed as the capacity   to ``remember'' a certain amount of samples back in time. In other words, RNNs can retain information about the past, enabling it to discover temporal correlations between events that are far away from each other in the data.

Early models of RNNs suffered from both \textit{exploding and vanishing gradient} problems~\cite{pascanu2013difficulty}. As   advanced architectures of RNNs, \textit{Long Short-Term Memory} (LSTM) and \textit{Gated Recurrent Unit} (GRU) were designed to  successfully address  the gradient problems and emerged among the other architectures. 

In this work, we profitably exploit the LSTM and GRU models in our  prediction framework.  
Furthermore, we also employ the \textit{bidirectional} version of both LSTM and GRU, dubbed BiLSTM and BiGRU, respectively, which differ   from the unidirectional ones since they connect two hidden layers of opposite directions to the same output;  in this way, the output layer can get information from past (backward) and future (forward) states simultaneously. 



%
{\bf Setting.\ }
 In both prediction tasks, we used the same configuration with  2 layers of 256 cells each with \textit{tanh} activation function, under a Keras 
  implementation on a Tensorflow backend.\footnote{\url{https://keras.io/}, \  \url{https://www.tensorflow.org/}}      
 To prevent overfitting, we introduced 
  both regular and recurrent \textit{dropout}~\cite{recurrent_dropout}.  In this regard, we   tested different values for regular and recurrent dropout and found that the use of smaller values of recurrent dropout, w.r.t. the regular one, can lead to slightly better results. However, given the negligible nature of the performance improvement,   we utilized the same value for both types of dropout, ranging between 30 and 60\%. 
  In addition, we leveraged the   \textit{batch normalization}~\cite{batch_norm} technique, with  batch size equal to 32. 
    Moreover, each of our RNN models was trained using the ADAM~\cite{adam} optimization algorithm   with   start-learning rate set to 0.002. This is a computationally efficient technique for gradient-based optimization of stochastic objective functions, which has shown to be particularly useful when dealing with large datasets or high-dimensional parameter space. 
 Finally,  we set 100  training epochs for both the prediction tasks. 




\subsection{Preprocessing}
We designed three steps of preprocessing of the ICBHI sound data: \textit{frame composition}, \textit{feature extraction}, and \textit{feature normalization}. We elaborate on each of these steps next.

\subsubsection{Frame composition}
In the first step of our preprocessing scheme, we segment  every respiratory cycle based on a sliding window of variable size, as described in Table~\ref{tab:setting}. Subsequently, for each portion (i.e., window) of the respiratory cycle, we   extract the Mel-Frequency Cepstral Coefficients (MFCCs)  (cf. Sect.~\ref{par:mfccs}) and finally concatenate the coefficients of each window. The resulting group of cepstral features constitutes a \textit{frame}, which represents the basic unit of data fed into the recurrent neural network.


As shown in Table~\ref{tab:setting}, we devised 7 configurations by varying the size of the window, the step between consecutive windows, 
  and the number of windows concatenated together after the extraction of the MFCCs.  
 Note that the settings S1, S3, S4, and S6 are characterized by   window size and window   step  of equal size, which  results in a null overlap of two consecutive windows, and produces non-overlapping partitioning of the whole respiratory cycle. 
Conversely, the remaining settings  correspond to a window step of half the size of the window, resulting in a 50\% overlap between consecutive windows.

\begin{table}[t!]
\centering
\caption{Configurations for the generation of RNN input frames from respiratory cycles}
\label{tab:setting}
\scalebox{0.9}{
\begin{tabular}{|c|c|c|c|c|c|}
\hline 
\multicolumn{1}{|p{1.2cm}|}{\centering \textbf{Setting id}} & \multicolumn{1}{|p{1.7cm}|}{\centering \textbf{Window size\\ $[ms]$}} & \multicolumn{1}{|p{1.7cm}|}{\centering \textbf{Window step\\ $[ms]$}} & \multicolumn{1}{|p{1.8cm}|}{\centering \textbf{\#windows}} & \multicolumn{1}{|p{1.7cm}|}{\centering \textbf{Frame size\\ $[ms]$}} & \multicolumn{1}{|p{1.8cm}|}{\centering \textbf{\#features}} \\ \hline \hline 
S1 & 500   & 500  & 1  & 500 & 13  \\ \hline
S2 & 500   & 250  & 1  & 500 & 13  \\ \hline
S3 & 250   & 250  & 1  & 250 & 13  \\ \hline
S4 & 50    & 50   & 5  & 250 & 65  \\ \hline
S5 & 50    & 25   & 5  & 150 & 65  \\ \hline
S6 & 50    & 50   & 10 & 500 & 130 \\ \hline
S7 & 50    & 25   & 10 & 275 & 130 \\ \hline
\end{tabular}
}
\end{table}

\subsubsection{Feature extraction}
\label{par:mfccs} 
For the extraction of significant features, we rely on Mel-Frequency Cepstral Coefficients (MFCCs)~\cite{mfcc}. 
In speech recognition, MFCC model has been widely and successfully used thanks to its ability in representing   the speech amplitude spectrum in a compact form. 

In our framework, the extraction of   MFCCs starts by dividing the input signal into frames of equal length and then applying a window function, such as the Hamming window to reduce spectral leakage. Next, for each frame, we generate a cepstral feature vector and apply the direct Fourier transform (DFT). While   information about the phase of the signal is discarded, the amplitude spectrum is retained and subject to logarithmic transformation, in order to  mimic the way   the human brain perceives the loudness of a sound~\cite{young2002htk}. 
 Moreover, to smooth the spectrum and emphasize perceptually meaningful  frequencies, we  aggregate the spectral components into a lower number of frequency bins. Finally, we apply   the discrete cosine transform (DCT)  to decorrelate the filter bank coefficients and yield a compressed representation.

\subsubsection{Feature normalization}
\label{sec:normalization}
Normalizing the input to a neural network is known to make training faster by limiting the chances of getting stuck in local minima (i.e.,  faster approaching to global minima at error surface)~\cite{normalization}. 
Within this view, we leverage two classic normalization techniques, 
Min-Max normalization and Z-score normalization (i.e., standardization). 
Recall that Z-score transformation of a feature value is calculated by subtracting  the population mean by it and dividing this difference by the population standard deviation.  Observed values above the mean have positive standard scores, while values below the mean have negative standard scores. 
 By contrast, Min-Max normalization (i.e., subtracting the minimum of all values from each specific one and dividing the difference by the difference between maximum and minimum)  scales feature values to a fixed range    [0,1]. 
\subsection{Evaluation and assessment criteria}
\label{subsec:evaluation}

For both  prediction tasks under consideration, 
we divided the ICBHI dataset into   80\% for training and 20\% for testing. 
We used  two  groups of assessment criteria: i)   ICBHI-specific  criteria, based     on \textit{micro-averaging}, as required by the ICBHI Challenge, and ii) \textit{macro-averaging} based criteria.  
The former group includes \textit{sensitivity} and  \textit{specificity}, and their average, named \textit{ICBHI-score}. 
Following  the   procedure described in~\cite{data,jakovljevic2018hidden}:   
$$\textit{Sensitivity} =  \frac{C_{crackles\_or\_wheezes}}{N_{crackles\_or\_wheezes}},$$ for the 2-class testbed,  
$$\textit{Sensitivity} =  \frac{C_{crackles} + C_{wheezes} + C_{both}}{(N_{crackles} + N_{wheezes} + N_{both}},$$  for the 4-class testbed, and $$\textit{Specificity}   = \frac{C_{normal}}{N_{normal}},$$ 
%
%
 %
 where $C$s and $N$s values denote   the number of correctly recognized instances  and the total number of instances, respectively, that belong to the  class   $crackles$, $wheezes$,  $both$ (resp. $crackles\_or\_wheezes$), in the 4-class (resp. 2-class) testbed, or $normal$. 
 Analogous definitions follow  for the evaluation of pathology-driven prediction; for instance, in the 3-class testbed:  
 $$\textit{Sensitivity}   =  \frac{C_{chronic} + C_{non\mbox{-}chronic}}{N_{chronic} + N_{non\mbox{-}chronic}}$$ 
 $$\textit{Specificity}   = \frac{C_{healthy}}{N_{healthy}}.$$
%
  
  We also considered macro-averaged \textit{accuracy}, \textit{precision}, \textit{recall} (sensitivity), and \textit{F1-score}, i.e., each of such scores is obtained as the average score over all classes. For instance, the 3-class pathology-driven evaluation accuracy is defined as: 
  $$\textit{Accuracy}=\frac{1}{3} \bigg(\frac{C_{chronic}}{N_{chronic}} + \frac{C_{non\mbox{-}chronic}}{N_{non\mbox{-}chronic}} + \frac{C_{healthy}}{N_{healthy}}\bigg).$$

\vspace{2mm}
\section{Experimental Results}

{\bf Plan of experiments and goals.\ }
 We organize the presentation of experimental results into four sections, which correspond  to our main goals of evaluation. 
First, we investigated the   impact of feature normalization   on the prediction performance of our framework (Sect.~\ref{subsec:norm}).  
Second, we  compared  the different types of RNNs considered in our framework, i.e., LSTM and GRU models, in their unidirectional and bidirectional architectures (Sect.~\ref{sec:results:RNNmodelscomparison}).     
 Third, we comparatively evaluated  our approach to    other  methods in the context of the ICBHI Challenge, i.e., for the  anomaly-driven prediction task (Sect.~\ref{subsec:icbhi}), and fourth, we  conducted an analogous evaluation stage for   the pathology-driven prediction task (Sect.~\ref{subsec:diseases}). 




\subsection{Impact of feature normalization on RNN performance}
\label{subsec:norm}
We analyzed whether and to what extent normalization of the MFCC features is beneficial for the prediction performance of our framework.  
Table~\ref{tab:normalization} reports accuracy results corresponding to  the LSTM model, for various frame-composition settings,  in the anomaly-driven prediction task, for both the   binary testbed (i.e., presence/ absence of anomalies) and four-class testbed (i.e., normal, presence of crackles, presence of wheezes, presence of both anomalies). 
 
Looking at the table, there is a clear evidence that   the use of Z-score normalization  generally leads to higher prediction accuracy, with significant improvements w.r.t. both min-max normalization and non-normalization of the features. This particularly holds for    the four-class testbed. 
%
 %
 
 The above finding was also confirmed by the other types of RNN used in our framework, with relative differences across the settings that revealed to be very similar to those observed for the LSTM model.  For this reason, in the following we will present results corresponding to   Z-score normalized  features. 

\begin{table}[t!]
\caption{Accuracy performance by LSTM models in the anomaly-driven prediction task, for the binary and four-class testbeds.}
\label{tab:normalization}
\scalebox{0.8}{
\begin{tabular}{|l|c|c||c|c||c|c|}
\hline
\multicolumn{1}{|c|}{\multirow{2}{*}{\textbf{Method}}} & \multicolumn{2}{c||}{\textbf{Un-normalized data}} & \multicolumn{2}{c||}{\textbf{Min-Max Normalization}} & \multicolumn{2}{c|}{\textbf{Z-score Normalization}} \\ \cline{2-7} 
\multicolumn{1}{|c|}{} & 2-Class & 4-Class & 2-Class & 4-Class & 2-Class & 4-Class \\ \hline \hline
LSTM-S1 & 0.74 & 0.69 & 0.68 & 0.64 & 0.78 & 0.72 \\ \hline
LSTM-S2 & 0.75 & 0.67 & 0.68 & 0.68 & 0.77 & 0.73 \\ \hline
LSTM-S3 & 0.75 & 0.69 & 0.73 & 0.68 & \textbf{0.81} & \textbf{0.74} \\ \hline
LSTM-S4 & 0.76 & \textbf{0.70} & 0.77 & \textbf{0.73} & 0.79 & \textbf{0.74} \\ \hline
LSTM-S5 & 0.77 & 0.69 & \textbf{0.79} & 0.72 & 0.79 & 0.72 \\ \hline
LSTM-S6 & \textbf{0.78} & 0.68 & 0.77 & 0.70 & 0.77 & 0.73 \\ \hline
LSTM-S7 & 0.76 & \textbf{0.70} & 0.79 & 0.72 & 0.80 & 0.72 \\ \hline
\end{tabular}
}
\end{table}

\subsection{Comparison of RNN models}
\label{sec:results:RNNmodelscomparison}

\begin{figure}[t!]
    \centering
    \includegraphics[width=0.8\linewidth]{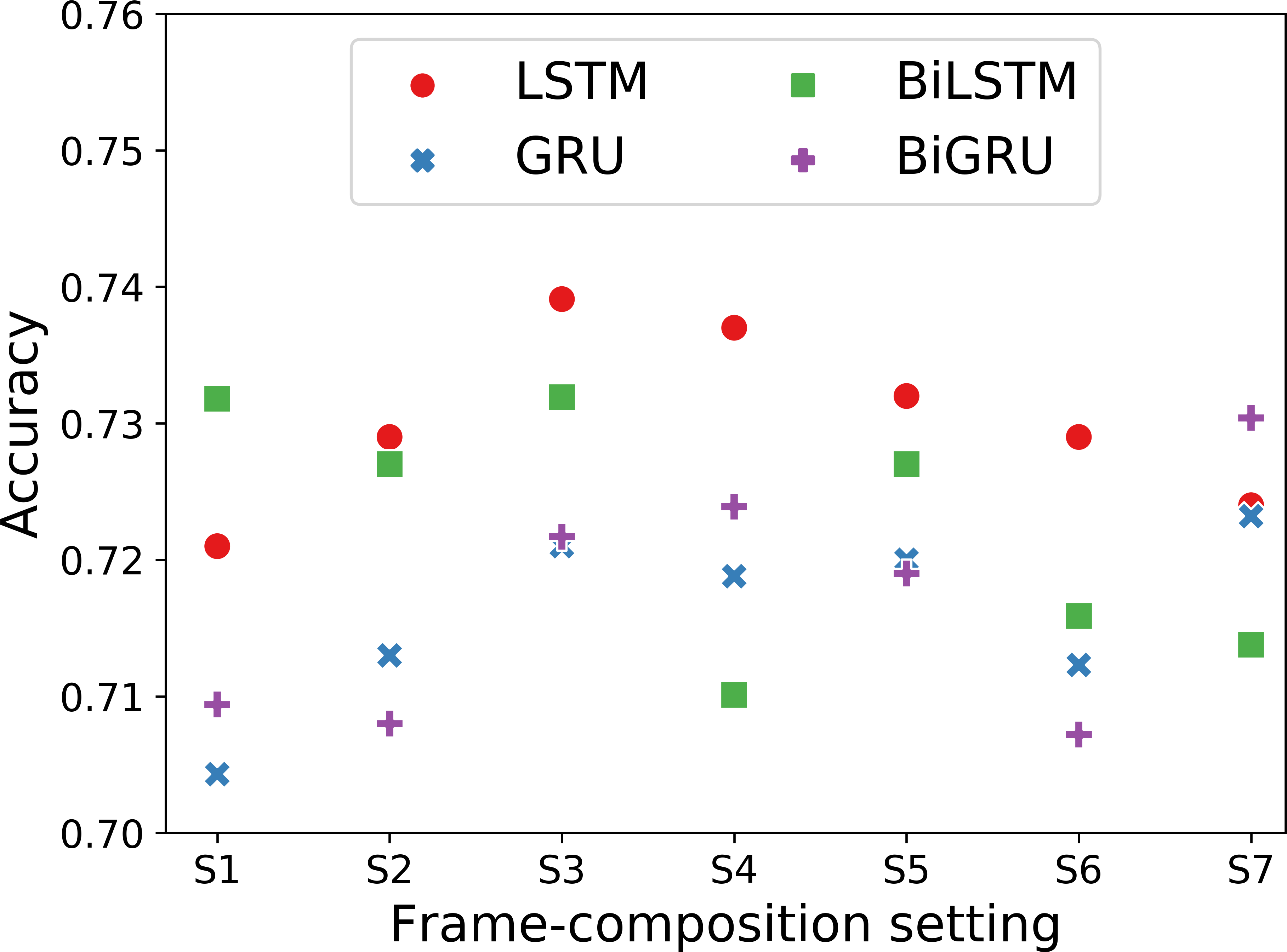}
    \caption{Comparison of RNN models in   four-class anomaly-driven prediction} 
    \label{fig:rnn_arch_comp}
\end{figure}

Figure~\ref{fig:rnn_arch_comp} shows the accuracy  obtained by the four different types of RNN models considered in our framework, i.e., LSTM, GRU, BiLSTM and BiGRU, for all frame-composition settings described in  Table~\ref{tab:setting}.

We observe that   all  architectures lead to relatively close   performance, ranging between 0.70 and 0.74 across the different settings. Overall, the largest differences correspond to settings S4 and S1, whereby  the  BiLSTM model behaves alternately as the worst and the best solution, respectively. Also, the unidirectional   GRU model tends to perform worse than the other models.  In general, the LSTM models provide consistently better results in most cases, though at the expense of memory and training efficiency; in this regard,  using the binary anomaly-driven prediction   as a case in point, the time required to complete the training composed of  100 epochs was about 13 minutes for LSTM, 11 minutes for GRU, 26 minutes for BiLSTM, and 22 minutes for BiGRU.\footnote{Experiments were carried out on a GNU/Linux (Mint 18) machine with Intel i7-3960X CPU and 64 GB RAM.} 
Due to space limitations, in the following we will present results obtained by the use of the LSTM model in our framework.

\subsection{Comparison with the  ICBHI Challenge competitors}
\label{subsec:icbhi}
We compared our  approach to 
 methods that participated to the ICBHI Challenge (Sect.~\ref{sec:related}). In addition, we also included the CNN-based method in~\cite{perna2018convolutional}, which was not previously tested on  the anomaly-driven prediction task. 
 
 %

Results    in Table~\ref{tab:icbhi_challenge} indicate that 
 our LSTM models clearly outperfom  all the competitors in terms of all three criteria. 
 Note that the frame-composition settings that correspond  to   the best  ICBHI-score in the challenge (i.e.,  73\%)  are S2, S3 and S4, which are characterized by a different frame-size (i.e.,  500, 250, and 50~ms), with  total number of MFCCs equal to 13, 13, and 65, respectively. It should be noted that  the relative difference in terms of  ICBHI-score w.r.t. the other frame-composition settings is just 1-2\%, which indicates      robustness of our LSTM-based framework to a crucial step in the preprocessing of respiratory sound data.

\begin{table}[t!]
\centering
\caption{ICBHI Challenge results on the detection of crackles and wheezes (four-class anomaly-driven prediction)}
\label{tab:icbhi_challenge}
\scalebox{0.835}{
\begin{tabular}{|c|c|c|c|}
\hline
\textbf{Method} & \textbf{Specificity} & \textbf{Sensitivity} & \textbf{ICBHI Score} \\ \hline \hline
Boosted Tree~\cite{chambres2018automatic} & 0.78 & 0.21 & 0.49 \\ \hline
CNN~\cite{perna2018convolutional} &  0.77  &  0.45 & 0.61 \\ \hline
HNN~\cite{berouti1979enhancement} &   $na$    & $na$  & 0.39 \\ \hline
MNRNN~\cite{kochetov2018noise} & 0.74 & 0.56 & 0.65 \\ \hline
STFT+Wavelet~\cite{serbes2018automated} & 0.83 & 0.55 & 0.69 \\ \hline
\hline
LSTM-S1 & 0.81 & 0.62 & 0.71 \\ \hline
LSTM-S2 & 0.82 & \textbf{0.64} & {0.73} \\ \hline
LSTM-S3 & {0.84} & \textbf{0.64} & \textbf{0.74} \\ \hline
LSTM-S4 & 0.83 & \textbf{0.64} & {0.73} \\ \hline
LSTM-S5 & 0.81 & 0.62 & 0.71 \\ \hline
LSTM-S6 & {0.84} & 0.60 & 0.72 \\ \hline
LSTM-S7 & \textbf{0.85} & 0.62 & \textbf{0.74} \\ \hline
\end{tabular}
}
\end{table}



\subsection{Performance on the pathology-driven prediction tasks}
\label{subsec:diseases}
Table~\ref{tab:cnn_rnn} summarizes  performance results obtained by our LSTM-based framework against the CNN-based competitor~\cite{perna2018convolutional}  on the pathology-driven prediction task, in both binary  (i.e., $healthy$ or $unhealthy$) and ternary (i.e., $healthy$, $chronic$, or $non\mbox{-}chronic$ diseases) fashion. 

Looking at the  results for the binary testbed, the best overall performance is achieved by our LSTM-based methods, in particular with frame-composition settings S4 and S7, which allow us to outperform the CNN-based method with gains up to   16\% accuracy, 9\% recall, 6\% F1-score, 4\% specificity, 3\% sensitivity, and 3\% ICBHI-score.  
 The ternary testbed results strengthen the superiority of the LSTM-based methods vs. the CNN-based one, in all cases. Again, settings S7 and S4 lead to the best performance of our methods, which should be ascribed by the beneficial effect due to higher number of features and finer-grain windowing used to generate the RNN input frames.

\begin{table}[t!]
    \centering
    \caption{Performance of our LSTM-based methods vs. CNN-based method, in the   pathology-driven classification tasks.}
    \label{tab:cnn_rnn}
    \scalebox{0.7}{
    \begin{tabular}{|c|c|c|c|c|c|c|c|c|c|}
    \hline
    \textbf{\#classes}\!\!&\!\!\textbf{Method} & \textbf{Accuracy} & \textbf{Precision} & \textbf{Recall} & \textbf{F1-score} & \textbf{Specif.} & \textbf{Sensitiv.} & \textbf{ICBHI}\\
     & & & & & & & & \textbf{score} \\
    \hline \hline
    2 & CNN~\cite{perna2018convolutional} &  0.83   & \textbf{0.95} &  0.83  &   0.88  & 0.78 & 0.97 & 0.88\\ \hline
    2 & LSTM-S1 & 0.98 & 0.92 & 0.85 & 0.88 & 0.70 & \textbf{1.00} & 0.85\\ \hline
    2 & LSTM-S3 & 0.98 & 0.93 & 0.87 & 0.89 & 0.77 & 0.99 & 0.88\\ \hline
    2 & LSTM-S4 & \textbf{0.99} &  \textbf{0.95}  & \textbf{0.92} & \textbf{0.94} & 0.79 & \textbf{1.00} & 0.89\\ \hline
    2 & LSTM-S6 & 0.98 & 0.92 & 0.88 & 0.90 & 0.80 & 0.99 & 0.90\\ \hline
    2 & LSTM-S7 & \textbf{0.99} & 0.94 & 0.91 &  0.92  & \textbf{0.82} & 0.99 & \textbf{0.91}\\ \hline
    \hline
    3 & CNN~\cite{perna2018convolutional} &  0.82  & 0.87 & 0.82 & 0.84 & 0.76 & 0.89 & 0.83\\ \hline
    3 & LSTM-S1 & 0.97 & 0.91 & 0.88 & 0.89 & 0.75 & 0.97 & 0.86\\ \hline
    3 & LSTM-S3 & 0.97 & 0.92 & 0.88 & 0.90 & 0.80 & \textbf{0.98} & 0.89\\ \hline
    3 & LSTM-S4 & \textbf{0.98} & 0.91 & \textbf{0.90} & 0.90 & 0.80 & \textbf{0.98} & 0.89\\ \hline
    3 & LSTM-S6 & 0.97 & 0.91 & 0.87 & 0.89 & \textbf{0.82} & \textbf{0.98} & \textbf{0.90}\\ \hline
    3 & LSTM-S7 & \textbf{0.98} & \textbf{0.93} & \textbf{0.90} & \textbf{0.91} & \textbf{0.82} & \textbf{0.98} & \textbf{0.90}\\ \hline
    \end{tabular}
    }
\end{table}

\section{Conclusion and Future Work} 
In this work, we developed a novel  deep-learning framework that originally integrates MFCC-based preprocessing of sound data and advanced Recurrent Neural Network models   for the detection of respiratory abnormal sounds (crackles and wheezes) and of   chronic/non-chronic diseases. 
Our empirical findings, drawn from an extensive evaluation conducted on the ICBHI Challenge data and against different competitors, suggest that our RNN-based framework advances the state-of-the-art in two respiratory disease prediction tasks, i.e., at anomaly-level and pathology-level. 

Our pointers for future research include the use or mixing of alternative    DL architectures, and an investigation of the impact of alternative representation models for the respiratory sounds on the prediction performance of our framework. In particular, we are interested in developing  hybrid models that can take advantage from a combination of time-series representation, whether in time or frequency domain, and  MFCCs.   



\end{document}